\documentclass[journal]{IEEEtran}
\usepackage{cite,graphicx,amssymb,amsmath,color,textcomp}
\usepackage{extarrows,multirow,multicol}
\begin{document}
\title{Outage Analysis of  Ambient Backscatter Communication Systems}
\author{ Wenjing Zhao,  Gongpu Wang, Saman Atapattu, Chintha Tellambura, and Hao Guan
\thanks{W. Zhao and G. Wang are with School of Computer and Information Technology, Beijing
Jiaotong University, Beijing 100044, China (e-mail: wenjing\_bjtu@163.com; gpwang@bjtu.edu.cn).}
\thanks{S. Atapattu are with the
Department of Electrical and Electronic Engineering,
The University of Melbourne, Melbourne, VIC 3010, Australia (e-mail: saman.atapattu@unimelb.edu.au).}
\thanks{C. Tellambura is with the Department of Electrical and Computer Engineering,
University of Alberta, Edmonton, AB T6G 2V4, Canada (e-mail: chintha@
ece.ualberta.ca).}
\thanks{H. Guan is with Nokia Bell Labs, Beijing, China (e-mail:hao.guan@nokia-sbell.com).}
}
\maketitle
\begin{abstract}
This paper addresses the problem of outage characterization of an ambient backscatter communication system  with a pair of passive tag and reader.
In particular, an exact expression for the effective channel distribution is derived. Then, the outage probability at  the reader is  analyzed rigorously.
Since the expression contains an infinite sum,  a tight truncation  error bound has been derived  to facilitate  precise numerical evaluations. Furthermore, an asymptotic expression is provided for high signal-to-noise ratio (SNR) regime.
\end{abstract}


\section{Introduction}\label{s_intro}
%
Backscatter communication systems, such as  Radio
Frequency Identification (RFID) system,  enable connecting
massive small computing devices  specially
for applications in Internet of Things (IoT) \cite{al2015st}.
The traditional RFID system  typically consists of  a
  tag and a reader. The reader first generates and transmits
an electromagnetic wave signal to the tag, and then the tag
receives and backscatters the signal to the reader.

One disadvantage  for the RFID system is that
the reader  needs  an oscillator to  transmit  a carrier signal,
for which dedicated encoding/decoding circuitry and power supply are required \cite{lee1998book}.
While these are essential components for a successful communication,
such in-built technology may no longer be promising
for small-scale devices. To overcome such overheads, ambient
backscatter prototypes are proposed in \cite{liu2013sigcomm, parks2014sigcomm}.



 The ambient backscatter technology utilizes environmental wireless signals (e.g., digital
 TV broadcasting, cellular or Wi-Fi) for both energy harvesting and information transmission,
which avoids battery as well as manual maintenance.
Specifically, the tag indicates bit 1 or bit 0
through  reflecting or non-reflecting state,
 and  the reader decodes the
received backscattered signal accordingly \cite{wang2016tcom}.
Ambient backscatter  may be widely used for future applications (e.g., many
applications in IoT with sensors located in dangerous spots
filled with poisonous gases/liquids, or inside building walls)
that are inconvenient and unsafe for wired communications
\cite{kuester2013mmw}.


The performance analysis of the ambient backscatter communication
is considered over real Gaussian channels in \cite{zhang2017spawc},
and complex Gaussian channels in \cite{xing2017wcnc, qian2017tcom}.
The  bit error rate (BER)  is derived
and the BER-based outage probability is   obtained  in \cite{zhang2017spawc}
for ambient bakcscatter communication systems
with energy detector.
In addition, the outage capacity
optimization problem is investigated in \cite{xing2017wcnc} when successive
interference cancellation (SIC) method is applied.
Besides, the
BER-based outage probability of a semi-coherent detection
scheme is calculated in \cite{qian2017tcom} in the case of perfect and imperfect
channel state information (CSI), respectively.

To our best knowledge,   effective channel distribution for ambient backscatter communication systems has not be addressed   and   the  outage performance based on  signal-to-noise ratio (SNR)  remains  an open problem, which is the focus of this paper.

In this paper, we consider an ambient backscatter communication system over real Gaussian channels. We  derive  an exact expression for the effective channel distribution in this system.  Particularly, we evaluate the outage performance  and analyse its asymptotic outage performance at high transmit SNR. Moreover,
 since the derived outage probability is the summation of infinite items, the corresponding truncation error bound is calculated.

\section{System Model}\label{s_sys}
\begin{figure}
  \centering
  \includegraphics[height=50mm,width=85mm]{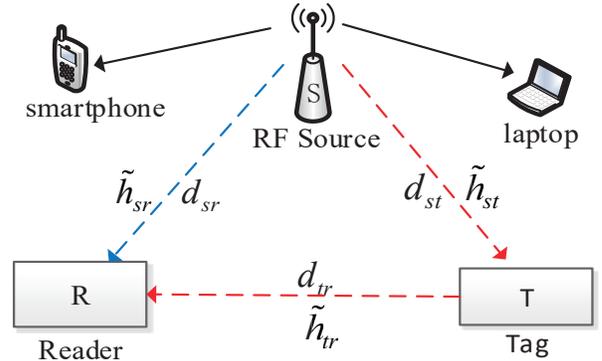}\\
  \caption{An ambient backscatter communication system.}\label{SystemModel}
\end{figure}

We consider an ambient backscatter communication system comprised of  an ambient RF source ($S$) and a pair of passive tag ($T$)  and  reader ($R$) (Fig.~\ref{SystemModel}).
While the RF source communicates with its legacy users (e.g., smartphones, laptops, etc.), both tag and reader may also receive that source signal. 
The tag first harvests energy from the source signal,
and then communicates with the reader via ambient backscatter.
Particularly,  the tag can backscatter or consume the energy of the received signal 
 to represent two states ``1" or ``0" for the reader, respectively \cite{wang2016tcom}.

The fading channels
of $S-R$, $S-T$, and $T-R$ links are denoted as  $\tilde h_{sr}$, $\tilde h_{st}$ and $\tilde h_{tr}$, respectively, which are real Gaussian random variables (RVs) distributed as $\tilde{h}_{sr}\sim \mathcal{N}(0,\tilde\sigma_{sr}^2)$, $\tilde{h}_{st}\sim \mathcal{N}(0,\tilde\sigma_{st}^2)$ and $\tilde{h}_{tr}\sim \mathcal{N}(0,\tilde\sigma_{tr}^2)$, where $\tilde\sigma_{sr}^2$, $\tilde\sigma_{st}^2$ and $\tilde\sigma_{tr}^2$ are channel variances. Further, the corresponding distances are $d_{sr}$, $d_{st}$ and $d_{tr}$, respectively.
Without loss of generality, we consider time instance $n$. The signal  received at the tag can be given as
\begin{equation}\label{SignalReceivedAtTag}
y_t(n)=\frac{\tilde h_{st}}{\sqrt{d_{st}^\alpha}} s(n),
\end{equation}
where $s(n)$ is the source signal with the average power $P$,
and $\alpha$ is the path-loss exponent.
The signal backscattered by the tag can be written as
\begin{equation}\label{SignalBackscatteredByTag}
x(n)=\eta B(n) y_t(n),
\end{equation}
where $B(n)\in \{0,1\}$ is a binary symbol and $\eta \in [0,1]$ is the attenuation factor.
Then, the received signal at the reader can be given as \cite{wang2016tcom} 
\begin{equation}\label{e_rxreaderor}
\begin{split}
y_r(n)&=\frac{\tilde h_{sr}}{\sqrt{d_{sr}^\alpha}} s(n) + \frac{\tilde h_{tr}}{\sqrt{d_{tr}^\alpha}} x(n)+ w(n) \\
&=h s(n) + w(n),
\end{split}
\end{equation}
where $w(n)$ is the additive white Gaussian noise (AWGN) at the reader with zero mean and $\sigma_w^2$  variance, and $h$ is the effective channel gain which can be given for two states as
\begin{align}\label{ModelExpression}
h&=\left\{
    \begin{array}{ll}
     h_{sr},&B(n)=0, \\
     h_{sr}+\eta h_{st} h_{tr}, &B(n)=1,
    \end{array}
  \right.
\end{align}
where $h_{sr}\triangleq \frac{\tilde{h}_{sr}}{\sqrt{d_{sr}^\alpha}}\sim \mathcal{N}(0,\sigma_{sr}^2)$,
$h_{st}\triangleq \frac{\tilde{h}_{st}}{\sqrt{d_{st}^\alpha}}\sim \mathcal{N}(0,\sigma_{st}^2)$, and $h_{tr}\triangleq \frac{\tilde{h}_{tr}}{\sqrt{d_{tr}^\alpha}}\sim \mathcal{N}(0,\sigma_{tr}^2)$ with $\sigma_{sr}^2=\frac{\tilde\sigma_{sr}^2}{d_{sr}^\alpha}$, $\sigma_{st}^2=\frac{\tilde\sigma_{st}^2}{d_{st}^\alpha}$ and $\sigma_{tr}^2=\frac{\tilde\sigma_{tr}^2}{d_{tr}^\alpha}$.
%
\section{Performance Analysis}\label{s_per}

\subsection{Effective Channel and SNR Distributions}\label{ss_chpdf}
We first derive the probability density function (PDF) of the effective channel, $f_h(x)$, which can be given as
\begin{align}\label{ChannelPDF}
f_h(x)=\left\{
    \begin{array}{ll}
     \frac{1}{\sqrt{2 \pi \sigma^2_{sr}}} e^{-\frac{x^2}{2 \sigma^2_{sr}}},\hspace{1.95cm} B(n)=0,& \\
     \frac{e^{\frac{\sigma^2_{sr}}{4 \eta^2 \sigma^2_{st} \sigma^2_{tr}}}}{\sqrt{2 \pi^3 \sigma^2_{sr} }} e^{-\frac{x^2}{2\sigma_{sr}^2}} \displaystyle\sum_{k=0}^{\infty} \frac{2^k \Gamma^2\left(k+\frac{1}{2}\right)x^{2k}}{(2k)!\sigma_{sr}^{2k}}& \\
     \hspace{0.3cm}\times W_{-k,0}\left(\frac{\sigma^2_{sr}}{2 \eta^2 \sigma^2_{st} \sigma^2_{tr}}\right), \hspace{0.65cm} B(n)=1,&
    \end{array}
  \right.
\end{align}
where $W_{a,b}(\cdot)$  and $\Gamma(\cdot)$ are  the Whittaker function \cite[eq.~(9.223)]{gradshteyn2007book} and the Gamma function \cite[eq.~(8.310.1)]{gradshteyn2007book}, respectively.
The proof is in Appendix~\ref{ss_proof1}.

The receive SNR at the reader is 
$\rho=\frac{P h^2}{\sigma^2_w} = \bar{\rho}h^2$
where $\bar{\rho}=\frac{P}{\sigma^2_w}$ may be the average transmit SNR. With a variable transformation for (\ref{ChannelPDF}), we can derive the PDF of $\rho$,
$f_\rho(x)$, as 
\begin{equation}
\begin{split}
f_\rho(x)&=\frac{\left[f_{h}\left(\sqrt{\frac{ x}{\bar{\rho}}}\right)+f_{h}\left(-\sqrt{\frac{x}{\bar{\rho}}}\right)\right]}{2\sqrt{\bar{\rho} x}} =\frac{f_{h}\left(\sqrt{\frac{x}{\bar{\rho}}}\right)}{\sqrt{\bar{\rho}  x}},
\end{split}
\end{equation}
where the second equality follows as 
the PDF $f_{h}(x)$ is an even function, i.e., $f_{h}(x)=f_{h}(-x)$.


\subsection{Outage Probability}\label{ss_out}

The outage probability is the probability that the SNR at the reader falls below a certain predetermined threshold $\rho_t$. Thus, it can be derived as
\begin{align}
P_{o}&= \Pr[\rho \leq \rho_t]=\int_0^{\rho_t}f_\rho(x)dx  \notag \\
&=\left\{
    \begin{array}{ll}
     \textrm{erf}\left(\sqrt{\frac{\rho_t}{ 2\bar\rho \sigma^2_{sr} }}\right), &\hspace{-1.2cm}B(n)=0, \\
     \displaystyle\sum_{k=0}^{\infty} \frac{ 2^{2k} \Gamma^2(k+\frac{1}{2})}{\sqrt{\pi^3}(2k)!} W_{-k,0}\left(\frac{\sigma_{sr}^2}{2 \eta^2 \sigma_{st}^2 \sigma_{tr}^2}\right)   \\
     \hspace{-0.05cm}\times e^{\frac{\sigma_{sr}^2}{4 \eta^2 \sigma_{st}^2  \sigma_{tr}^2}}\gamma\left(k+\frac{1}{2},\frac{\rho_t}{ 2\bar\rho \sigma^2_{sr} } \right), &\hspace{-1.2cm}B(n)=1,
    \end{array}
  \right. \label{OutageProbability}
\end{align}
where $\textrm{erf}(\cdot)$ and  $\gamma(\cdot,\cdot)$  are the  error function \cite[eq.~(8.250.1)]{gradshteyn2007book} and the incomplete gamma function \cite[eq.~(8.350.1)]{gradshteyn2007book}, respectively.
The equation \eqref{OutageProbability} can be obtained by following from \cite[eq.~(3.321.2)]{gradshteyn2007book}, \cite[eq.~(8.250.1)]{gradshteyn2007book} and \cite[eq.~(3.381.1)]{gradshteyn2007book}.

\subsection{Truncation Error Bound}\label{ss_err}

Since the outage probability expression for $B(n)=1$ case in \eqref{OutageProbability}  is with an infinite sum, it is a challenge for numerical calculation.
We therefore truncate it into a finite number of terms $T$ in order to ensure a given numerical accuracy requirement. Then, we bound the truncation error as
\begin{equation}
\label{e_errbound}
\begin{split}
|\epsilon_T| &\leq \frac{ \Psi\left(\frac{1}{2},0,\nu\right)}{\sqrt{\pi \nu} T!} \biggr[\sqrt{\frac{2  \rho_t }{\sigma^2_{sr} \bar{\rho} }} \gamma\left(T+1,\frac{  \rho_t }{2 \sigma^2_{sr} \bar{\rho}}\right)  \\
&\, \,\, \,\, \,\, \, \, \,\, \,\, \,\, \,\, \, \, \, \,\, \,\, \,\, \,\, \, \, \,\,\, \,\, \,\, \,\,~~~~~~~-2 \gamma\left(T+\frac{3}{2},\frac{ \rho_t }{2 \sigma^2_{sr} \bar{\rho} }\right)\biggr],
\end{split}
\end{equation}
where $\nu=\frac{\sigma_{sr}^2}{2 \eta^2 \sigma_{st}^2 \sigma_{tr}^2}$, and $\Psi(\cdot,\cdot,\cdot)$ is the confluent hypergeometric function \cite[eq.~(9.211.4)]{gradshteyn2007book}. The proof is in Appendix~\ref{ss_proof2}.

\subsection{Asymptotic Analysis for High SNR}\label{ss_highsnr}

To further investigate the ambient backscatter system, we approximate outage of the reader for large SNR as
\begin{align}
P_{o}\approx\left\{
    \begin{array}{ll}
     \sqrt{\frac{2 \rho_t}{\pi \sigma_{sr}^2}}\frac{1}{\sqrt{\bar\rho}},&\hspace{-0.3cm}B(n)=0, \\
     \frac{\sqrt{2\rho_t} e^{\frac{\sigma_{sr}^2}{4 \eta^2 \sigma_{st}^2 \sigma_{tr}^2}}  W_{0,0}\left(\frac{\sigma_{sr}^2}{2 \eta^2 \sigma_{st}^2 \sigma_{tr}^2}\right)}{\sqrt{\pi \sigma_{sr}^2}} \frac{1}{\sqrt{\bar\rho}}, &\hspace{-0.3cm}B(n)=1.
    \end{array}
  \right. \label{e_highsnr}
\end{align}
The proof is in Appendix~\ref{ss_proof3}.

Interestingly, when transmit SNR tends to infinity, we get that the diversity gain is
$-\underset {\textrm{SNR}\rightarrow \infty} {\lim}\frac{\log P_{\textrm{out}}}{\log \textrm{SNR}}= \frac{1}{2}$ for both cases $B(n)=0$ and $B(n)=1$.
Accordingly, when power is big enough, the outage probability of the reader is inversely proportional to  square root of the power.

\section{Numerical and Simulation  Results}

\begin{figure}
  \centering
  \includegraphics[height=65mm,width=93mm]{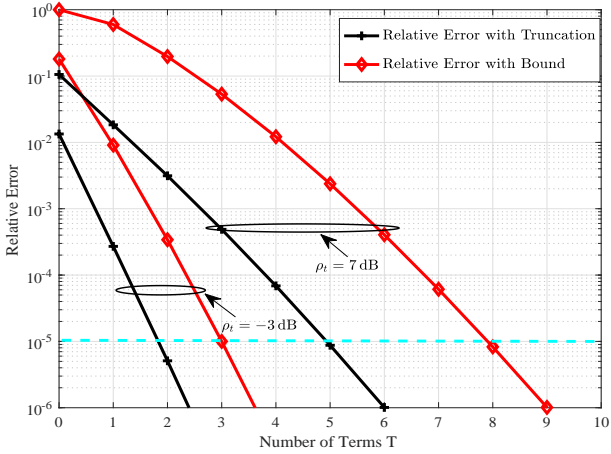}\\
  \caption{The relative error versus the number of terms $T$ for $\rho_t=-3$dB and $\rho_t=7$dB when $\bar{\rho}=3$dB and $B(n)=1$.}\label{error}
\end{figure}

This section provides simulation results based on the system model in Section \uppercase\expandafter{\romannumeral2}
and numerical results based on the analytical results in Section \uppercase\expandafter{\romannumeral3}.
Here, the attenuation factor $\eta$ is  set as 0.7.
Besides, we assume $\sigma_{sr}^2=\sigma_{st}^2=1$ and $\sigma_{tr}^2=3$, unless otherwise specified.


For the state $B(n) = 1$, we derive a truncation error bound \eqref{e_errbound}.
Fig.~\ref{error} shows the relative error versus the number of terms $T$ when $\bar{\rho}=3$dB for $\rho_t=-3$dB and $\rho_t=7$dB.
We calculate the relative error with truncation as $\frac{|\textrm{exact}-\textrm{truncated}|}{\textrm{exact}}$;
the relative error with bound as $\frac{\textrm{error bound in} \, \eqref{e_errbound}}{\textrm{exact value}}$; the exact value with
numerical integration; and the truncated value for different $T$ by using (\ref{OutageProbability}).
The relative error with truncation is less than $10^{-5}$ when $T>2$ and $T>5$ for $\rho_t=-3$dB and $\rho_t=7$dB, respectively.
The relative error with bound is less than $10^{-5}$ when $T>3$ and $T>8$ for $\rho_t=-3$dB and $\rho_t=7$dB, respectively.
This shows the tightness of the bound.
Moreover, by observation, we may say that a very accurate value can be calculated using small $T$.

\begin{figure}
  \centering
  \includegraphics[height=65mm,width=93mm]{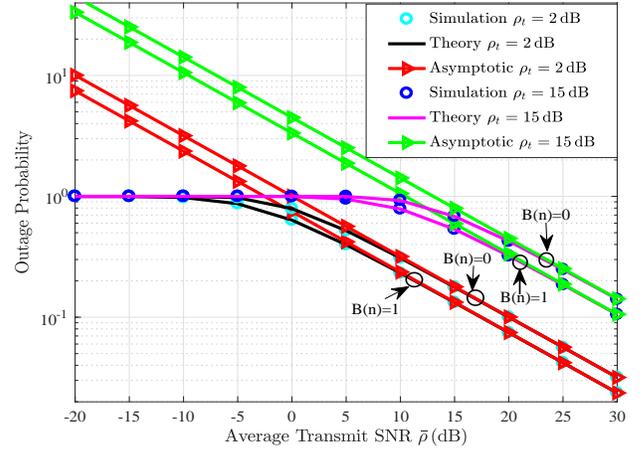}\\
  \caption{The outage probability $P_o$ versus the average transmit SNR $\bar{\rho}$.}\label{Pout_SNR}
\end{figure}

Fig.~\ref{Pout_SNR} illustrates the outage probability $P_o$ versus the average transmit SNR $\bar{\rho}$  when  $\rho_{t}=2$dB and $\rho_{t}=15$dB. The outage probability decreases with the increasing average transmit SNR. The asymptotic expressions (\ref{e_highsnr}) also approach the exact values asymptotically  at high SNR. Since the diversity gain is 1/2, the slope of the asymptotic outage curves is 1/2.

\begin{figure}
  \centering
  \includegraphics[height=65mm,width=93mm]{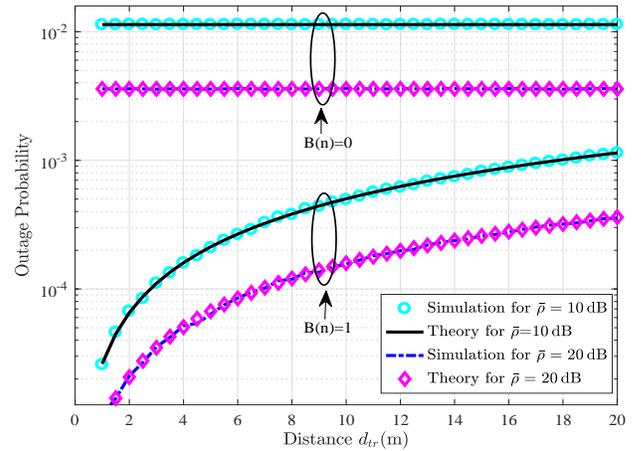}\\
  \caption{The outage probabilities $P_o$ versus the distance $d_{tr}$ between the tag and the reader.}\label{DISTANCE}
\end{figure}

 Fig.~\ref{DISTANCE} depicts the  outage probabilities $P_o$ versus the distance $d_{tr}$ between the tag and the reader. We consider both cases of SNR $\bar{\rho}=10$dB and $\bar{\rho}=20$dB, respectively. We set $\rho_{t}=2\,$dB, $\alpha=3$, $d_{sr}=20$ meters and $d_{st}=20$ meters. Besides, the channel variances  $\tilde{\sigma}_{sr}^2$, $\tilde{\sigma}_{st}^2$ and $\tilde{\sigma}_{tr}^2$ are set as 1, 1 and 3, respectively.
 The outage probability $P_o$ is a constant in the case of $B(n) = 0$ due to no transmission  between the tag
and the reader.
However, when enlarging the distance $d_{tr}$ between the tag and the reader, the outage probability $P_o$ witnesses a upward trend in the case of $B(n)=1$.
For example, if we expect $P_o<10^{-4}$, the distance $d_{tr}$ between the tag and the reader should not exceed 3 meters when $\bar{\rho}=10$dB and 7 meters when $\bar{\rho}=20$dB.

\section{Conclusion}
Ambient backscatter, a new form of wireless communication, has potential commercial value as well as a series of open problems.
In this paper, we first derived  the effective channel distribution for the ambient backscatter communication system.
We next analyzed the outage probabilities,
its truncation error bound  as well as the asymptotic outage probabilities at  high  SNR.
It was found that the asymptotic outage probabilities could  well approach the exact values,
and our truncation error bound could provide a reasonable estimation of the truncation terms.

\appendix

\subsection{Proof of (\ref{ChannelPDF})}\label{ss_proof1}
The distribution of any real Gaussian channel $h_{ab}\in\{h_{sr},h_{st},h_{rt}\}$ is
\begin{align}
\label{e_hpdf}
f_{h_{ab}}(x)=\frac{1}{\sqrt{2 \pi \sigma^2_{ab}}}e^{-\frac{x^2}{2 \sigma^2_{ab} }},
\end{align}
where $\sigma^2_{ab}\in\{\sigma^2_{sr},\sigma^2_{st},\sigma^2_{rt}\}$.

In the case of $B(n)=1$, we have $h=h_{sr}+\eta h_{st} h_{tr}$.
The distribution $f_\xi(x)$ of $\xi=\eta h_{st}h_{tr}$ can be shown as \cite{simon2006book}
\begin{equation}
\label{e_xipdf}
f_\xi(x)=\frac{1}{\pi \eta \sqrt{\sigma^2_{st}\sigma^2_{tr}}} K_0\left(\frac{  |x| }{\eta \sqrt{\sigma^2_{st}\sigma^2_{tr}}}\right),
\end{equation}
where $K_\alpha(\cdot)$ is the modified Bessel function of the second kind \cite{gradshteyn2007book}.

Since the two random variables $h_{sr}$ and $\xi$ are independent,  the distribution $f_{h}(x)$ is the convolution of $f_{h_{sr}}(x)$ and $f_\xi(x)$. Therefore, we can obtain
\begin{align}
f_{h}(x)
&= \int_{-\infty}^{\infty}f_{h_{sr}}(x-z) f_\xi(z)\,dz \notag \\
&\overset{\text{(a)}}{=} \frac{ e^{-\frac{x^2}{2 \sigma_{sr}^2}}}{\delta \varphi}
\int_{0}^{\infty} K_0(\frac{z}{\varphi}) e^{-\frac{z^2}{2 \sigma_{sr}^2}} \left(
e^{\frac{ x z}{\sigma_{sr}^2}} + e^{-\frac{x z}{\sigma_{sr}^2}}\right) d z  \notag \\
&\overset{\text{(b)}}{=} \frac{2e^{-\frac{x^2}{2 \sigma_{sr}^2}}}{\delta \varphi}\int_{0}^{\infty} K_0(\frac{  z }{\varphi}) e^{-\frac{z^2}{2 \sigma_{sr}^2}}\cosh(\frac{ x z}{\sigma_{sr}^2})
d z \notag \\
&\overset{\text{(c)}}{=} \frac{2e^{-\frac{x^2}{2 \sigma_{sr}^2}}}{\delta \varphi} \sum_{k=0}^{\infty}
\frac{x^{2k}}{\sigma_{sr}^{4k}(2k)!} \int_{0}^{\infty} K_0(\frac{  z }{\varphi}) e^{-\frac{z^2}{2 \sigma_{sr}^2}} z^{2k} d z \notag \\
&\overset{\text{(d)}}{=} \frac{e^{\frac{\nu}{2}}}{\delta} \sum_{k=0}^{\infty}\frac{2^k \Gamma^2(k+\frac{1}{2})W_{-k,0}(\nu)}{(2k)! \sigma_{sr}^{2k}}e^{-\frac{x^2}{2\sigma_{sr}^2}}x^{2k},
\label{total}
\end{align}
where
\begin{align}
\delta=\sqrt{2 \pi^3 \sigma^2_{sr}}, \quad \varphi=\eta \sqrt{\sigma^2_{st} \sigma^2_{tr} }, \quad \nu=\frac{\sigma_{sr}^2}{2 \eta^2 \sigma_{st}^2 \sigma_{tr}^2},  \label{proofeq1}
\end{align}
 (a) follows by two integrals having  identical bounds,
 (b) follows by using the definition of hyperbolic cosine $\cosh x=\frac{1}{2}(e^x+e^{-x})$,
 (c) follows by replacing $\cosh x$ with its series expression $\sum_{k=0}^{\infty} \frac{x^{2k}}{(2k)!}$,
 and (d) follows from \cite[eq.~(6.631.3)]{gradshteyn2007book}.

\subsection{Proof of \eqref{e_errbound}}\label{ss_proof2}
On the basis of \eqref{OutageProbability}, the truncation error $\epsilon(T)$ with the number of terms $T$ can be bounded as
\begin{align}
&\epsilon(T) \notag \\
=&\frac{e^{\frac{\nu}{2}}}{\sqrt{\pi}\pi} \sum_{k=T+1}^{\infty} \frac{ 2^{2k} \Gamma^2\left(k+\frac{1}{2}\right)W_{-k,0}(\nu)}{(2k)!} \gamma\left(k+\frac{1}{2},\frac{ \rho_t}{2 \sigma^2_{sr} \bar{\rho} }\right) \notag \\
\overset{\text{(a)}}{=}& \frac{\sqrt{\nu}}{\pi} \sum_{k=T+1}^{\infty}  \frac{1}{k!}\int_0^{\infty}\frac{e^{-\nu x} x^{k+\frac{1}{2}}}{(1+x)^{k+\frac{1}{2}}}dx \int_0^{\frac{\rho_t}{2 \sigma^2_{sr} \bar{\rho}}}e^{-y}y^{k-\frac{1}{2}}dy   \notag \\
\overset{\text{(b)}}{=}& \frac{\sqrt{\nu}}{\pi (T+1)!}   \int_0^{\infty} \int_0^{\frac{\rho_t}{2 \sigma^2_{sr} \bar{\rho}}} \frac{e^{-\nu x} x^{T+\frac{3}{2}}}{(1+x)^{T+\frac{3}{2}}}
 e^{-y}y^{T+\frac{1}{2}}   \notag \\
&~~~~~~~~~~~~~~~~~~~~~~~~~~~\times  {_1F_1}\left(1;T+2;\frac{xy}{x+1}\right) dydx  \notag \\
\overset{\text{(c)}}{=}& \frac{\sqrt{\nu}}{\pi T!}   \int_0^{\infty} \int_0^{\frac{ \rho_t}{2 \sigma^2_{sr} \bar{\rho} }} \frac{e^{-\nu x} x^{\frac{1}{2}}}{(1+x)^{\frac{1}{2}}}
 e^{-y}y^{-\frac{1}{2}}   \notag \\
&~~~~~~~~~~~~~~~~~~~~~~~~~~~\times e^{\frac{xy}{x+1}} \gamma\left(T+1,\frac{xy}{x+1}\right) dydx  \notag 
\end{align}
\begin{align}
\mathop{<}\limits ^{(d)}&\frac{\sqrt{\nu}}{\pi T!}   \int_0^{\infty}  \frac{e^{-\nu x} x^{\frac{1}{2}}}{(1+x)^{\frac{1}{2}}}dx
 \int_0^{\frac{\rho_t }{2 \sigma^2_{sr} \bar{\rho} }}\frac{\gamma(T+1,y)}{\sqrt{y}} dy  \notag \\
=& \frac{ \Psi(\frac{1}{2},0,\nu)}{\sqrt{\pi \nu} T!} \biggr[\sqrt{\frac{2  \rho_t }{\sigma^2_{sr} \bar{\rho} }} \gamma\left(T+1,\frac{  \rho_t }{2 \sigma^2_{sr} \bar{\rho}}\right)  \notag \\
&\, \,\, \,\, \,\, \, \, \,\, \,\, \,\, \,\, \, \, \, \,\, \,\, \,\, \,\, \, \, \,\,\, \,\, \,\, \,\,~~~~~~~-2 \gamma\left(T+\frac{3}{2},\frac{ \rho_t }{2 \sigma^2_{sr} \bar{\rho} }\right)\biggr],   \notag \label{TruncationError}
\end{align}
where $\nu$ is defined in \eqref{proofeq1}, (a) is obtained  from \cite[eq.~(9.222.1)]{gradshteyn2007book} and \cite[eq.~(8.350.1)]{gradshteyn2007book}, (b) follows by setting $j=k-T-1$  and leveraging the integral representation of Hypergeometric function \cite[eq.~(9.211.4)]{gradshteyn2007book}, (c) utilizes the following  equation
\begin{align}
_1F_1 \left(1;T+2;\frac{xy}{x+1}\right)=(T+1)\left(\frac{xy}{x+1}\right)^{-(T+1)}e^{\frac{xy}{x+1}} \notag \\
\times \gamma\left(T+1, \frac{xy}{x+1}\right),\notag
\end{align}
(d) is based on $\frac{x}{x+1}<1$ for $x>0$, and $\Psi(\cdot,\cdot,\cdot)$ is the confluent hypergeometric function \cite[eq.~(9.211.4)]{gradshteyn2007book}.

\subsection{Proof of \eqref{e_highsnr}}\label{ss_proof3}

In the case of $B(n)=0$ and using series representation of the error function $\textrm{erf}(\cdot)$ \cite[eq.~(8.253.1)]{gradshteyn2007book}, we can expand \eqref{OutageProbability}  as
\begin{align}
 \textrm{erf}\left(\sqrt{\frac{\rho_t}{ 2\bar\rho \sigma^2_{sr} }}\right)=\frac{2}{\sqrt{\pi}} \displaystyle \sum_{k=0}^\infty \frac{(-1)^k {\rho_t}^{k+1/2} }{k! (2k+1) (2\bar\rho \sigma^2_{sr})^{k+1/2}}. \notag
\end{align}
For $\bar\rho \rightarrow \infty$,  we consider the lowest exponent for $1/\bar\rho$,
i.e., the index $k=0$.
Similarly, in the case of $B(n)=1$, we can expand $\gamma\left(\cdot,\cdot\right)$ in \eqref{OutageProbability} as
\begin{align}
\gamma\left(a,x\right)=\sum _{j=0}^{\infty} \frac{(-1)^j x^{a+j}}{j! (a+j)}. \notag
\end{align}
We then consider the lowest exponent for $1/\bar\rho$, i.e.,  the indies $k=0$ and $j=0$. Thereby,
when average transmit SNR  tends to infinity, namely, $\bar\rho\rightarrow\infty$, the asymptotic outage probability  can be simplified as \eqref{e_highsnr}.



\end{document}